\documentclass[11pt]{article}
\usepackage{amsfonts}
\usepackage{amsmath}
\usepackage{amsthm}
\usepackage{graphicx}
\usepackage{rotating}
\usepackage{array, color}
\usepackage{times}
\usepackage{bm}
\usepackage{natbib}
\usepackage{graphicx}
\usepackage{rotating}
\usepackage{amsfonts}
\usepackage{array, color}

\def\({{\Bigl(}}
\def\){{\Bigr)}}
\def\[{{\Bigl[}}
\def\]{{\Bigr]}}
\def\LARE{{\rm LARE}}
\def\LPRE{{\rm LPRE}}
\def\GRE{{\rm GRE}}

\makeatletter\@addtoreset{equation}{section}
\renewcommand\theequation
{\ifnum \c@section>\z@ \thesection.\fi\@arabic\c@equation}
\makeatother

\begin{document}

\noindent{\Large \bf \textsf{Least Product Relative Error
Estimation}}

\vspace*{0.2in}

\noindent {\large \textsf{Kani C{\normalsize{\textsf{HEN}}},
Yuanyuan L{\normalsize{\textsf{IN}}}, Zhanfeng
W{\normalsize{\textsf{ANG}}}} and Zhiliang Y{\normalsize{\textsf{ING}}}}

\vspace*{0.2in} \footnotetext[1]{Kani Chen is Professor, Department of Mathematics, Hong Kong University of Science and Technology, Kowloon, Hong Kong, China
(E-mail:makchen@ust.hk).
Yuanyuan Lin is Assistant Professor,
Department of Statistics, School of Economics and Wang Yanan Institute for Studies in Economics, Xiamen University, China
(E-mail:linyy@xmu.edu.cn).
Zhanfeng Wang is Associate Professor,
Department of Statistics and Finance, University of Science and Technology of China, Hefei 230026, China (E-mail:
zfw@ustc.edu.cn).
Zhiliang Ying is Professor, Department of Statistics, Columbia
University, New York NY 10027 (E-mail: zying@stat.columbia.edu).
 }

\vspace{0.1in}

\noindent

A least product relative error criterion is proposed for
multiplicative regression models.
It is invariant under scale transformation
of the outcome and covariates. In addition, the objective function
is smooth and convex, resulting in a simple and uniquely defined
estimator of the regression parameter.   It is
shown that the estimator is asymptotically normal and that the
simple plugging-in variance estimation is valid. Simulation
  results confirm that the proposed method performs
well.  An application to body fat calculation is presented to
illustrate the new method.

\vspace{0.2in}

\section{Introduction}

In regression analysis, the least squares (LS) and least absolute deviation (LAD)  are the most
commonly used criteria based on absolute errors (Stigler, 1981;
Portnoy \& Koenker, 1997). In some situations, however, criteria based on relative
errors that are scale invariant and less sensitive to outliers
 are more desirable (Narula \& Wellington,
1977; Makridakis et al., 1984; Khoshgoftaar et al., 1992; Ye,
2007; Park \& Stefanski, 1998; Chen et al., 2010; Zhang \& Wang, 2012).
Consider the following multiplicative
regression model
\begin{eqnarray}
\label{1-1} Y_i=\exp({X}_i^\top {{\beta}})\epsilon_i,
\hskip 1cm i=1,\cdots,n,
\end{eqnarray}
where $Y_i$ is the response variable, ${X}_i$ is the $p$-vector
of explanatory variables with the first component being 1
(intercept), ${{\beta}}$ is the corresponding $p$-vector of
regression parameters with the first component being the intercept
and $\epsilon_i$ is the  error term, which is strictly
positive. An additional constraint on $\epsilon$ needs to be
imposed so that the first component of $\beta$ (intercept) becomes
identifiable. Model (\ref{1-1}) is also known as the accelerated
failure time (AFT) model in the survival analysis literature.

For the multiplicative regression model (\ref{1-1}), Chen {  et
al.} (2010) gives a convincing argument  that  a proper criterion
should take into account both types of relative errors: one
relative to the response and the other relative to the predictor
of the response. A criterion with only one type of relative errors
often leads to biased estimation. They introduce  the   least
absolute relative error  (LARE) estimation for model (\ref{1-1})
by minimizing
\begin{eqnarray}
\label{1-2} \LARE_n({{\beta}})\equiv \sum_{i=1}^n
\left\{\left|{Y_i-\exp({X}_i^\top {{\beta}})\over Y_i}\right|+
\left|{Y_i-\exp({X}_i^\top {{\beta}})\over \exp({X}_i^\top
{{\beta}})}\right|\right\},
\end{eqnarray}
the sum of the two types of the relative errors.
The LARE estimation enjoys the robustness and scale-free property. However,
like the LAD, the LARE criterion function is nonsmooth, and, as a result, the
limiting variance of the corresponding estimator involves the density of the error. Furthermore,
its computation is slightly more complicated than linear programming.

It would be desirable to develop a criterion function which not only incorporates the relative
error terms, but also is smooth and convex. The latter would ensure the numerical uniqueness of the
resulting estimator and the consistency of the usual plug-in sandwich-type variance estimation. The
main purpose of this paper is to introduce a simple, smooth,  convex and
interpretable criterion function
   and to develop a related inference procedure.

The rest of the paper is organized as follows. Section 2 introduces the least product relative error (LPRE)
criterion, along with simple inference procedures, including point and variance estimation, hypothesis testing
and related large sample properties. Extension of the LPRE to a general class of
relative error criteria is given in Section 3. Section 4 contains simulation results and a
real example.
Some discussion  and concluding remarks are given in Section 5.

\section{Least product relative error}

The least absolute relative error (LARE) criterion (\ref{1-2}) of Chen
{  et al.} (2010) is the result of adding together the two relative error
terms. In this paper, we consider multiplying the two relative error
terms and propose the following least product relative error
(LPRE) criterion
\begin{eqnarray}
\label{1-5} \LPRE_n({{\beta}})\equiv \sum_{i=1}^n \left\{
\left|{Y_i-\exp({X}_i^\top {{\beta}})\over Y_i}\right|\times
\left|{Y_i-\exp({X}_i^\top {{\beta}})\over
\exp({X}_i^\top {{\beta}})}\right|\right\}.
\end{eqnarray}
Note that the summand can be written as $\{Y_i-\exp({X}_i^\top
{{\beta}})\}^2/\{Y_i \exp({X}_i^\top {{\beta}})\}$. Thus,
it may be viewed as a symmetrized version of the squared relative
errors (Park and Stefanski, 1998).

A simple algebraic manipulation leads to the following alternative expression
\begin{eqnarray}
\label{1-6} \LPRE_n({{\beta}})\equiv \sum_{i=1}^n \left\{Y_i
\exp(-{X}_i^\top {{\beta}})+Y_i^{-1}\exp({X}_i^\top
{{\beta}})-2\right\},
\end{eqnarray}
from which we can see major advantages. First, the criterion function is infinitely differentiable. Second, it is strictly convex since the exponential function is strictly convex.  As a result, finding the minimizer is equivalent to finding the root of its first derivative. The usual asymptotic properties can therefore be derived by a local quadratic expansion and standard inference methods for M-estimation are applicable.

\subsection{Estimation}

We now deal with parameter estimation and develop the
corresponding theory. Our estimator for $\beta$ will be denoted by
${{\hat{\beta}}}_n$ and defined as the minimizer of (\ref{1-5})
or, equivalently, (\ref{1-6}). The strict convexity of (\ref{1-6})
entails that the minimizer, if it exists, must be unique. Assume
 the design matrix $\sum_{i=1}^n X_iX_i^\top$ is nonsingular. This is a
minimum condition for the purpose of identifiability. Then, $\LPRE_n({{\beta}})$
is strictly convex, and,
as
$\|\beta\|\to\infty$,
$\sum_{i=1}^n (X_i^\top \beta)^2\to\infty$, implying $\max \{ |X_i^\top
\beta|: i=1,..., n\} \to\infty$. It follows that
  $\LPRE_n({{\beta}})\to \infty$ as
$\|\beta\|\to\infty$. And the following theorem holds.

{\bf Theorem 1.} {\it If $\sum_{i=1}^n X_iX_i^\top$ is nonsingular,
then ${{\hat{\beta}}}_n$ exists and is unique.}

{\it Remark 1.} The nonsingularity of $\sum_{i=1}^n X_iX_i^\top$ is also
a necessary and sufficient condition for the least squares estimator
to be unique.

We next establish asymptotic properties for ${{\hat{\beta}}}_n$
under suitable regularity conditions. For notational simplicity,
we assume that $(X^\top, Y)^\top$, $(X_i^\top, Y_i)^\top$, $i=1,\dots, n$
are independent and identically distributed.
It allows for heteroskedasticity in that it does not
require the error term $\epsilon$ to be independent of the
explanatory variable $X$.  We will use the following conditions for
the development of the asymptotic theory.

Condition C1. There exists $\delta>0$ such that
$E\{(\epsilon+1/\epsilon)\exp{(\delta \|X\|)}\}<\infty$.

Condition C1*. There exists $\delta>0$ such that
$E\{(\epsilon+1/\epsilon)^2 \exp{(\delta \|X\|)}\}<\infty$.

Condition C2. The expected design matrix, $E(XX^\top)$, is
positive definite.

Condition C3.  The error terms satisfy $E(\epsilon|X)
=E(1/\epsilon|X)$.

Condition C1 is almost minimal for the criterion function (4) to have a finite expectation in a neighborhood of the true parameter $\beta_0$. It also ensures that the limit of (4)
is twice differentiable with respect to $\beta$ and that the
differentiation and expectation is interchangeable. Condition C2 ensures that the design matrix is nonsingular, a minimal requirement for the regression parameter to be identifiable. Under C1 and C2, the limiting criterion function is strictly convex in a neighborhood of $\beta_0$. Condition C3 is equivalent to that the derivative of the criterion function at $\beta_0$ has mean 0, again a minimal condition for the resulting estimator to be asymptotically unbiased. The strict convexity and the asymptotic unbiasedness ensure that the estimator is consistent. Condition C1* is simply a stronger version of C1 for the asymptotic normality to hold.

{\bf Theorem 2.} {\it Under Conditions C1, C2 and C3,
${{\hat{\beta}}}_n$ is strongly consistent.}

{\it Proof.} \ Under C1, C2 and C3, one can show that $\LPRE_n(\beta)/n$ converges to $E\{\LPRE_n(\beta)\}/n$  in a small neighborhood of $\beta_0$  and that both are convex. Thus, by Rockafellar (1970, Theorem 10.8), ${{\hat{\beta}}}_n$, the minimizer of $LPRE_n(\beta)$, converges to $\beta_0$, the minimizer of
$E\{LPRE_n(\beta)\}$.

The next theorem establishes the asymptotic normality and the validity of the plug-in variance
estimation. Let $D=E\{XX^\top(\epsilon+1/\epsilon)\}$ and
$V=E\{XX^\top(\epsilon-1/\epsilon)^2\}$. Define their
plug-in estimators
$\hat D=(1/n) \sum_{i=1}^n X_iX_i^\top  \{
{\exp({X_i^\top{{\hat{\beta}}}_n})}/{Y_i}+
{Y_i}/{\exp({X_i^\top{{\hat{\beta}}}_n})}\}$ and
$\hat V=(1/n) \sum_{i=1}^n X_iX_i^\top  \{
{Y_i}/{\exp({X_i^\top{{\hat{\beta}}}_n})}-
{\exp({X_i^\top{{\hat{\beta}}}_n})}/{Y_i}\}^2.$

{\bf Theorem 3.} {\it Under Conditions C1*, C2 and C3, $\sqrt n
({{\hat{\beta}}}_n-\beta_0)$ is asymptotically normal with mean
$0$ and covariance matrix $D^{-1}VD^{-1}$, which is consistently
estimated by $\hat D^{-1}\hat V \hat D^{-1}$. }

{\it Proof.} \ Since ${{\hat{\beta}}}_n$ is consistent, by the Taylor  expansion,
$${{\hat{\beta}}}_n-\beta_0=\hat D_*^{-1}\frac 1 n \sum_{i=1}^n X_i\left \{\frac
{Y_i}{\exp({X_i^\top{{{\beta}}_0}})}- \frac
{\exp({X_i^\top{{{\beta}}_0}})}{Y_i}\right\},$$
where $\hat D_*=(1/n) \sum_{i=1}^n X_iX_i^\top \{
{\exp({X_i^\top{{{\beta}}}^*})}/{Y_i}+
{Y_i}/{\exp({X_i^\top{{{\beta}}^*}})}\}$ and ${{\beta}}^*$
lies in between the true parameter $\beta_0$ and the LPRE estimate ${{\hat{\beta}}}_n$.
The desired results follow from the law of large numbers and the central limit theorem.

It can be shown that when the error $\epsilon$ has density
\begin{eqnarray}\label{eff-f}
f(x)=c\exp(-x-1/x-\log x+2)I(x>0),
\end{eqnarray}
where $c$ is the normalizing constant, $D=V$ becomes the Fisher information. It then follows that ${{\hat{\beta}}}_n$ is asymptotically efficient.

\subsection{Hypothesis testing}

We now turn to hypothesis testing. Although the asymptotic theory developed in the preceding subsection can be used to construct Wald-type testing statistics, we will focus on an approach that is based directly on the LPRE criterion. For simplicity, we assume homogeneous errors, i.e. $\epsilon$ is independent of $X$ and consider the following null hypothesis
{
\begin{eqnarray}\label{nullhyp}
H_0:~\beta\in \Omega_0=\{{b}\in R^{p}: \it{H}^\top{b}=0\},
\end{eqnarray}
where $\it{H}=(h_1,\cdots,h_q)$ and $h_j$, $j=1, \dots, q$ are $p$-vectors that are linearly
independent and lie in the linear space spanned by the column vectors of the design matrix $X$.

Let
 \begin{eqnarray}\label{Mn-lpre}
M_n\equiv \min_{{\beta}\in
\Omega_0} \LPRE_n({\beta})-\min_{{\beta}\in R^p} \LPRE_n({\beta}).
\end{eqnarray}
Through the usual quadratic expansion, we can arrive at an asymptotic ANOVA-type decomposition. The asymptotic normality can then be applied to show that, under the null hypothesis,
$M_n$ converges in distribution to $K\chi^2_q$, where $K=4E(\epsilon)/{E\{(\epsilon-1/\epsilon)^2\}}$ and $\chi^2_q$ is the central chi-squared distribution with $q$ degree of freedom. The constant $K$ can be estimated consistently by
$\hat K= 4\sum_{i=1}^n \{Y_i
\exp(-{X}_i^\top {{\hat{\beta}}_n})\}/\sum_{i=1}^n \{Y_i
\exp(-{X}_i^\top {{\hat{\beta}}_n})-
1/Y_i\exp({X}_i^\top {{\hat{\beta}}_n})\}^2$. Therefore we can use $M_n/\hat K$  as the testing statistic with $\chi^2_{q, 1-\alpha}$ as the cut-off point,
where $\alpha$ is a given nominal significance level.

\section{General relative error criteria}

A general relative error (GRE) criterion can be constructed:
\begin{eqnarray}
\label{1-3} \GRE_n({{\beta}})\equiv \sum_{i=1}^n g
\left(\left|{Y_i-\exp({X}_i^\top {{\beta}})\over Y_i}\right|,
\left|{Y_i-\exp({X}_i^\top {{\beta}})\over \exp({X}_i^\top
{{\beta}})}\right|\right),
\end{eqnarray}
where $g(a, b)$ is a bivariate function satisfying certain
regularity conditions. Taking $g(a,b)=a+b$, it becomes the LARE criterion function
(Chen {\it et al.}, 2010) while $g(a,b)=ab$,  it becomes the LPRE of the preceding section. One may also consider $g(a,b)=\max\{a, b\}$ (Ye, 2007). Note that all three criteria here are symmetric functions. A possible non-symmetric one could be
$g(a,b)=a+\exp(b)$, where we pay more attention to the relative error of $b$ and more
heavily penalize large value of $b$ compared to $a$.

The derivative of $\GRE_n$ with respect to $\beta$ is defined as
$$S_n(\beta)=\sum_{i=1}^n \phi
\left(\left|{Y_i-\exp({X}_i^\top {{\beta}})\over Y_i}\right|,
\left|{Y_i-\exp({X}_i^\top {{\beta}})\over \exp({X}_i^\top
{{\beta}})}\right|\right).$$
Its expectation becomes 0 when
\begin{eqnarray}\label{Gcon}
E\left(\phi
(\left|{1-\epsilon^{-1}}\right|,
\left|{\epsilon-1}\right|)|{X}\right)=0.
\end{eqnarray}
Let $\hat \beta _n$ be a minimizer of the criterion function (\ref{1-3}).
It follows that, under (\ref{Gcon}), $\hat\beta_n$ is asymptotically unbiased.
In fact we have the following result concerning the limiting distribution of $\sqrt n (\hat \beta _n-\beta_0)$.

{\bf Theorem 4.} {\it Under (\ref{Gcon}) and additional regularity conditions concerning the nonsingularity of the design matrix and finite moment condition on the error,
$\sqrt{n}({{\hat{\beta}}}_n-{{\beta}}_0) $ is asymptotically normal with mean $0$ and covariance matrix $(1/a^{2})J\it{V}^{-1}$ where $J=E[\{\phi(|\epsilon-1|,
|1-1/\epsilon|)\}^2]$, $\it{V}=E({XX}^\top)$ and constant $a$ satisfies,
as $ |c| \to 0$,
$$E\{\phi(|1-\epsilon\exp{(-c)}|,|\epsilon^{-1}\exp{(c)}-1|)\}=ac+o(|c|).\
\ $$ }

The proof of Theorem 4 is similar to that of Theorem 1 in Chen {\it et  al.} (2010) and is omitted.
In general, the asymptotic variance  of ${{\hat{\beta}}}_n$,
the minimizer of $\GRE_n({{\beta}})$, may involve the density
function of the error. To avoid density estimation,
 an approximation based on random weighting can be applied.
If the error
$\epsilon$ has a density function as follows:
\begin{eqnarray}\label{eff_f}
f(x)=c\exp\{-g(|1-x|,|1-x^{-1}|)-\log x\}I(x>0),
\end{eqnarray}
where $c$ is a normalizing constant,  then the estimator
${{\hat{\beta}}}_n$ is asymptotically efficient.
Density $f(x)$ in (\ref{eff_f})
belongs to a class of inverse transformation invariant density, meaning that
if a random variable $X$ is distributed with
density $f(x)$, then $1/X$ is equal in distribution
to $X$. Figure 1 shows densities of some particular choices of
function $g$.
One can see that the error distribution with which the product criterion is efficient has heavier tails than others, indicating that the product criterion is more robust
in practical application.

\begin{center}
INSERT FIGURE 1 HERE
\end{center}

Based on general relative error criterion (\ref{1-3}),
a general test statistic to test hypothesis (\ref{nullhyp}) 
can be constructed as
 \begin{eqnarray}\label{Mn}
M_n\equiv \min_{{\beta}\in
\Omega_0} \GRE_n({\beta})-\min_{{\beta}\in R^p} \GRE_n({\beta}).
\end{eqnarray}
Especially, when
the error terms follow the distribution described in (\ref{eff_f}),
$M_n$ is identical to the log-likelihood ratio test statistic.
The following theorem demonstrates the asymptotic distribution of $M_n$.

{\bf Theorem 5.} {\it Under the hypothesis (\ref{nullhyp}) and additional regularity conditions,
$$M_n \, {\rightarrow} \,  \frac{J}{2a}\chi_q^2 \quad \hbox{in distribution,}
$$
as $n\rightarrow \infty$,
where $\chi_q^2$ refers to the   chi-square distribution with $q$ degrees of freedom.
}

The proof of Theorem 5 is similar to that of Theorem  1 in Chen {\it et  al.} (2008) and is also omitted.
In general,  the asymptotic distribution of
$M_n$ may involve the density of the errors.
The plug-in method involving density estimation can be inaccurate and
computationally troublesome. In this case, a random weighting method, as used by
Chen {\it et  al.} (2008) and Wang {\it et  al.} (2009),
can be applied.

\section{Numerical studies}
\subsection{Simulation studies}
 Simulation studies are conducted to compare the finite sample
performance of the proposed LPRE, the LARE, the LS and the LAD. The
data are generated from the model
\begin{eqnarray}
\label{1-10} Y=\exp(\beta_0+\beta_1X_{1}
+\beta_2X_{2})\epsilon,\hskip 1cm
\end{eqnarray}
where $X_{1}$ and $X_{2}$ are two independent covariates
following the standard normal distribution $ N(0,1)$ and  $(\beta_0,~\beta_1,~\beta_2)^\top=(1,1,1)^\top$. We
consider five error distributions:  the
distribution with which the LARE estimator is efficient;
  the
distribution with which the LPRE estimator is efficient;
the exponential  of the uniform  distribution on $(-2,2)$;
 the log-standard normal distribution;   and  the
uniform distribution   on $(0.5,a)$
with $a$   chosen such that $E(\epsilon)=E(1/\epsilon)$. Note that
the first four error distributions are such  that $1/\epsilon$ is distributed same as
  $\epsilon$.
The sample size $n$ is $200$. The
variance estimation for the LARE and the LAD is based on random weighting with
resampling size $N=500$, while
the variance estimation for the LPRE and the LS is based on the plug-in rule.
The LS and LAD estimates are obtained by minimizing $\sum_{i=1}^n(\log
Y_i-\beta_0- \beta_1X_{1i} -\beta_2X_{2i})^2$ and
$\sum_{i=1}^n|\log Y_i-\beta_0- \beta_1X_{1i} -\beta_2X_{2i}|$,
respectively. The simulation results are based on 1000 replications.
\begin{center}
INSERT TABLES 1, 2, 3 HERE
\end{center}
It is seen from Table 1 that  the LPRE
performs considerably better than the LARE, the LS
and the LAD when $\log(\epsilon)$ is uniformly distributed on $ (-2,2)$. With log-normal
error distribution,
  the LPRE performs as well as
the LARE and is comparable to the LS. With the uniform error distribution,
 the  LPRE
performs slightly better than the LS, and much better than the LARE and the LAD.
The variance estimation of the LPRE  gives accurate
coverage probability in the study.

The performance of the proposed test statistic $M_n$ is evaluated
with the product relative error criterion.
We consider two null hypotheses $H_0: \beta_2=0$ and $ H_0: \beta_1=\beta_2=0$.
Tables 2 and 3 present the
empirical significance levels and powers with $n=200$ when $\epsilon$ follows the distributions with which the LPRE and the LARE are respectively efficient, the log-uniform distribution on $(-2, 2)$ and
the log-standard normal distribution.
 It is seen that the empirical significance levels
are close to the nominal levels, suggesting that $M_n$
is  adequate. Under $H_0: \beta_2=0$  and  nominal level  0.05,
the power increases from 0.05 to 1.0 as $\beta_2$ varies from 0.0 to 0.4.
In other words, the power increases as the parameters move away
from the null hypothesis, a common phenomenon in   hypothesis testing.

\subsection{Application}

We apply the proposed method to analysis of
the body fat data. The data contain  various body measurement indices related with
percentage of body fat for 252 men, which are available at
\emph{http://lib.stat.cmu.edu/datasets/ bodyfat}; see Penrose {\it et  al.} (1985).
   We select 12 explanatory variables: age ($X_1$), height$^4$/weight$^2$ ($X_2$) and 10 other
body circumference indexes (neck,
chest, abdomen, hip, thigh,
knee, ankle,  biceps,  forearm
and wrist, denoted by $X_i$, $i=3,\ldots, 12$). We note that $X_2$ is a transform of the well-known
body mass index (BMI) defined as the ratio of weight to height$^2$.
The sample size $n$  is 251.
The response variable $Y$ is the percentage of body fat. We delete one observation with
$Y=0$ and fit the model
\begin{eqnarray}
\label{5-2} Y_i=\exp(\beta_0+\sum_{j=1}^{12}\beta_jZ_j)
\epsilon_i,\hskip 1cm i=1,\cdots,n,
\end{eqnarray}
  where $Z_j,~j=1,\cdots,12$, denote
the normalized explanatory variables.

To evaluate the performance of different methods, the dataset is partitioned into
two parts. The first part with  200 observations is used to fit model (\ref{5-2}),  and
the rest 51 observations are used to evaluate the prediction power. The results are shown in Tables 4 and 5.
\begin{center}
INSERT TABLES 4 and 5 HERE
\end{center}
The $p$-value is calculated by  $1-\Phi(|{\hat{\beta}_j}/\hat s_j|)$, where ${\hat{\beta}_j}$ is the estimate of $\beta_j$,
$\hat s_j$ is the estimated standard deviation  for ${\hat \beta_j}$, and $\Phi(\cdot)$ is the   cumulative distribution function for the standard normal distribution.
The variance estimation of the LPRE and the LS are obtained by the plug-in rule, while that for LARE and LAD are obtained by random weighting resampling.
Table 4 shows that the four methods identify some common
variables (with $p$-value $<$ 0.05), such as age, 1/BMI and abdomen circumference.
It makes sense that  the
percentage of body fat increases as age, BMI and abdomen circumference become large.
However, the biceps circumference is identified only by
the LPRE, indicating the percentage of body fat
increases as the biceps circumference becomes larger.

The prediction accuracy based on the four methods estimation is measured by
four different median indices: median of absolute prediction errors $\{|Y_i-\hat Y_i|\}$ (MPE), median of
product relative prediction errors $\{|Y_i-\hat Y_i|^2/(Y_i\hat Y_i)\}$ (MPPE), median of
additive relative prediction errors $\{|Y_i-\hat Y_i|/Y_i+|Y_i-\hat Y_i|/\hat Y_i\}$ (MAPE) and median of squared prediction
errors $\{(Y_i-\hat Y_i)^2\}$ (MSPE), where $\hat Y_i=\exp(\hat\beta_0+\sum_{j=1}^{12}\hat\beta_jZ_j)$, $i=201,\cdots,251$.
Table 5 shows that the LPRE has the smallest
MPE, MPPE, MAPE and MSPE among the LPRE, the LARE, the LS and the LAD.

\section{Concluding remarks}
In our view, in the realm of criteria based on relative errors, the LPRE proposed
in this paper has the best potential to be the basic and primary choice, just like the
least squares in the realm of criteria based on absolution errors. The proposed LPRE represents a substantial  improvement  over that of  Chen {\it et al.} (2010) both theoretically
and computationally, particularly in terms of the simplicity of inference.
Extensions can be made to cover analysis of censored data and high dimensional data.
       Moveover, we present a more general
GRE method and    tests of linear hypotheses based on
relative errors, which is not studied in Chen {\it et al.} (2010) and other relevant literature.
The LPRE criterion may
have  broad  applications in financial
and survival data analysis.

 %
 %
\section*{References}
\bibliography{paper-ref}
{ \small
\begin{description}
\item
Chen, K., Guo, S., Lin, Y., and Ying, Z. (2010). Least absolute relative error estimation. \textit{J. Am. Statist. Assoc.} \textbf{105}, 1104--1112.

\item
Chen, K., Ying, Z., Zhang, H., and Zhao, L. (2008). Analysis of
least absolute deviation. \textit{Biometrika} \textbf{95}, 107--122.

\item
Gauss, C. F. (1809). \textit{Theoria Motus Corporum Coelestium}.
Perthes, Hamburg. Translation reprinted as \textit{Theory of the
Motions of the Heavenly Bodies Moving about the Sun in Conic
Sections}. Dover, New York, 1963.

\item
Khoshgoftaar, T. M., Bhattacharyya, B. B., and Richardson, G. D.
(1992). Predicting software errors, during development, using
nonlinear regression models: a comparative study. \textit{IEEE
Transactions on Reliability} \textbf{41}, 390--395.

\item
Knight, K. (1998). Limiting distribution for $L_1$ regression
estimators under general conditions. \textit{Ann. Statist.}
\textbf{26}, 755--770.

\item
Makridakis, S., Andersen, A., Carbone, R., Fildes, R., Hibon, M., Lewandowski, R.,
Newton, J., Parzen, E., and Winkler, R. (1984). {\it The Forecasting Accuracy of Major Time Series Methods}. New York: Wiley.

\item
Narula, S. C., and  Wellington, J. F. (1977). Prediction, linear
regression and the minimum sum of relative errors. \textit{Technometrics} \textbf{19}, 185--190.

\item
Park, H., and Stefanski, L. A. (1998). Relative-error prediction.
\textit{Statist.$\&$Prob. Letters} \textbf{40}, 227--236.

\item
Penrose, K. W., Nelson, A. G. and Fisher, A. G. (1985). Generalized body composition prediction equation for men using simple measurement techniques (abstract).
{\it Medicine and Science in Sports and Exercise} {\bf 17}, 189--189.

\item
Pollard, D. (1991). Asymptotics for least absolute deviations
regression estimators. \textit{Econometric Theory} \textbf{7},
186--199.

\item
Portnoy, S., and Koenker, R. (1997). The Gaussian hare and the
Laplacian tortoise: computability of squared-error versus
absolute-error estimators (with discussion). \textit{Statist. Sci.}
 \textbf{12}, 279--300.

\item
Rockafellar, R. T. (1970). {\it Convex analysis}. Princeton University Press, Princeton, N.J.


\item
Scheff$\acute{e}$, H. (1959). {\sl The analysis of variance}. New York: John Wiley.

\item
Stigler, S. M. (1981). Gauss and the invention of least squares.
\textit{Ann. Statist.} \textbf{9}, 465--474.

\item Wang, Z., Wu, Y. and Zhao, L. (2009). Approximation by randomly weighting method for linear hypothesis testing in censored regression model. \textit{Sci. in China Series A: Mathematics} \textbf{52}, 561--576.

\item
Ye, J. (2007). Price models and the value relevance of accounting
information. \textit{Technical report}.

\item Zhang, Q. and Wang, Q. (2012). Local least absolute relative error estimating approach for partially linear multiplicative model. \textit{Statist. Sinica} Preprint.

\end{description}
}

\begin{figure}[h!]
\begin{center}
\includegraphics[scale=0.8]{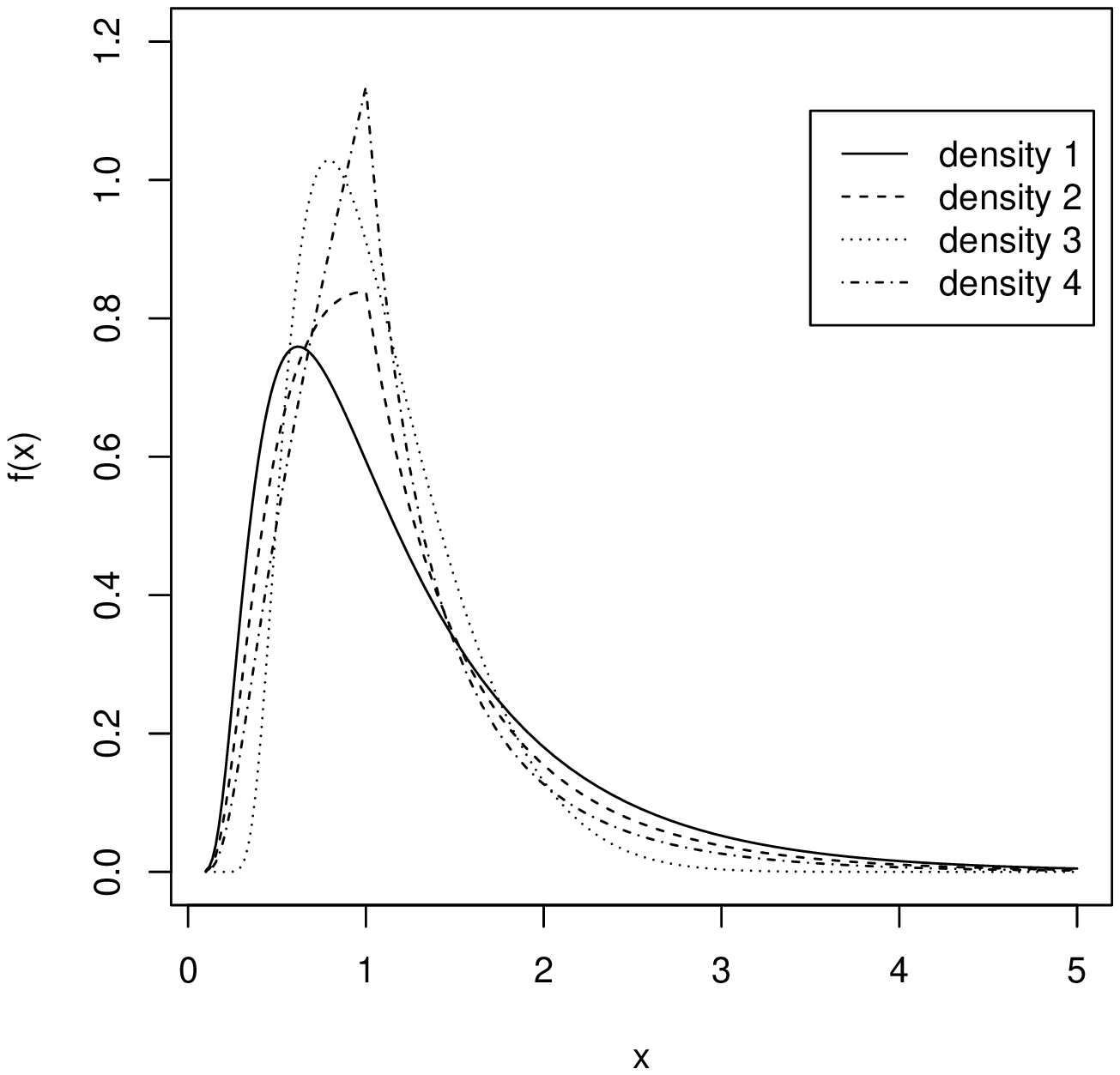}
\vskip .00021in
\vspace{1ex} \footnotesize{
\begin{eqnarray*}
 \mbox{density1:} && f_1(x)=c_1\exp(-x-x^{-1}-\log x+2)I(x>0);\\
\mbox{density2:} && f_2(x)=c_2\exp(-\max\{|1-x|,|1-x^{-1}|\}-\log x)I(x>0);\\
\mbox{density3:} && f_3(x)=c_3\exp\{-(1-x)^2-(1-x^{-1})^2-\log x\}I(x>0); \\
\mbox{density4:} && f_4(x)=c_4\exp(-|1-x|-|1-x^{-1}|-\log x)I(x>0).
\end{eqnarray*}
}\caption{Plot of four densities.} \label{p1}
\end{center}
\end{figure}

%
%
{
\begin{sidewaystable}[ht!]    
\tabcolsep=3pt \fontsize{9}{12}\selectfont
\vspace{0.1cm}
\begin{center}
\textbf{Table 1.} 
\textit{Comparison among various approaches with
${{\beta}}=(1,1,1)^\top$}
\end{center}
\begin{center} { 
\begin{tabular}{cc|c|c|c|c|c}
\hline  & & $\epsilon\sim f_1(\cdot)$ &
$\log(\epsilon)\sim Unif$(-2,2)& $ \log(\epsilon) \sim N(0,1)$& $\epsilon\sim f_2(\cdot)$& $\epsilon\sim Unif(0.5, a)$ \\
 \hline
 &  & $\hat{\beta}_0$ \hskip 0.3cm\  $\hat{\beta}_1$\hskip 0.3cm \  \hskip 0.3cm$\hat{\beta}_2$& $\hat{\beta}_0$\hskip 0.3cm \  $\hat{\beta}_1$\hskip 0.3cm
 \ \hskip 0.3cm $\hat{\beta}_2$ & $\hat{\beta}_0$ \hskip 0.3cm\  $\hat{\beta}_1$\hskip 0.3cm \
  \hskip 0.3cm $\hat{\beta}_2$&$\hat{\beta}_0$ \hskip 0.3cm\  $\hat{\beta}_1$\hskip 0.3cm \  \hskip 0.3cm$\hat{\beta}_2$ &$\hat{\beta}_0$ \hskip 0.3cm\  $\hat{\beta}_1$\hskip 0.3cm \  \hskip  0.3cm$\hat{\beta}_2$ \\
 \hline
 LPRE & BIAS & 0.007 \  0.004 \  0.009& 0.001 \  0.001
 \ 0.004 & 0.002 \  0.001 \  0.001 & 0.004 \  0.005 \  0.004& 0.002 \  0.000 \  0.001\\
& SE & 0.037 \  0.037 \  0.037& 0.067 \  0.067
 \ 0.067 & 0.074 \  0.075 \  0.076 & 0.045 \  0.045 \  0.045& 0.023 \  0.023 \  0.023  \\
& SEE & 0.036 \  0.036 \  0.036& 0.067 \  0.067
 \ 0.067 & 0.075 \  0.075 \  0.075 & 0.045 \  0.045 \  0.045& 0.023 \  0.023 \  0.023  \\
 & CP & 0.943 \  0.942 \  0.955& 0.952 \  0.952
 \ 0.952 & 0.948 \  0.949 \  0.945 & 0.945 \  0.956 \  0.947& 0.948 \  0.950 \  0.950 \\
 \hline
 LARE & BIAS & 0.001 \  0.002 \  0.001& 0.001 \  0.002
 \ 0.000 & 0.004 \  0.004 \  0.002& 0.001 \  0.001 \  0.001 & 0.044 \  0.000 \  0.001\\
 & SE & 0.032 \  0.033 \  0.034& 0.077 \  0.075
 \ 0.073 & 0.076 \  0.073 \  0.076 & 0.047 \  0.048 \  0.047 & 0.035 \  0.034 \  0.034\\
 & SEE & 0.033 \  0.034 \  0.034& 0.075 \  0.075
 \ 0.075 & 0.073 \  0.072 \  0.072 & 0.047 \  0.047 \  0.047 & 0.032 \  0.032 \  0.033 \\
& CP & 0.945 \  0.944 \  0.951& 0.944 \  0.943
 \ 0.959 & 0.926 \  0.928 \  0.931 & 0.936 \  0.927 \  0.934 & 0.942 \  0.943 \  0.943\\
\hline
 LS & BIAS & 0.001 \  0.002 \  0.001& 0.001 \  0.002
 \ 0.000 & 0.004 \  0.003 \  0.002&  0.004 \  0.005 \  0.005 & 0.004 \  0.005 \  0.001\\
 & SE & 0.035 \  0.035 \  0.037& 0.083 \  0.081
 \ 0.078 & 0.071 \  0.069 \  0.072 & 0.047 \  0.047 \  0.047  & 0.025 \  0.025 \  0.025\\
& SEE & 0.035 \  0.035 \  0.035& 0.081 \  0.080
 \ 0.080 & 0.070 \  0.069 \  0.070 & 0.045 \  0.045 \  0.045  & 0.025 \  0.026 \  0.026\\
& CP & 0.945 \  0.952 \  0.926& 0.948 \  0.937
 \ 0.951 & 0.950 \  0.939 \  0.935 &  0.939 \  0.941 \  0.950 & 0.945 \  0.946 \  0.947\\
 \hline
 LAD & BIAS & 0.001 \  0.002 \  0.001& 0.001 \  0.004
 \ 0.001 & 0.004 \  0.003 \  0.001 & 0.009 \  0.008 \  0.010 & {0.053} \  0.001 \  0.002\\
 & SE & 0.033 \  0.034 \  0.034& 0.143 \  0.140
 \ 0.135 & 0.090 \  0.085 \  0.090 & 0.061 \  0.060 \  0.058 & 0.037 \  0.037 \  0.036\\
 & SEE & 0.036 \  0.038 \  0.038& 0.145 \  0.144
 \ 0.144 & 0.093 \  0.094 \  0.094 & 0.063 \  0.062 \  0.062 & 0.040 \  0.040 \  0.039\\
& CP & 0.938 \  0.915 \  0.921& 0.897 \  0.868
 \ 0.888 & 0.917 \  0.907 \  0.906 &  0.899 \  0.887 \  0.906 & 0.900 \  0.902 \  0.901\\
 \hline
\multicolumn{6}{l}{{$f_1(x)=c_1\exp(-|1-x|-|1-x^{-1}|-\log x)I(x>0)$};}\\
\multicolumn{6}{l}{{$f_2(x)=c_2\exp(-x-x^{-1}-\log x+2)I(x>0)$}.}
\end{tabular}
}
\end{center}
\end{sidewaystable} 
}

\begin{table}[h!]
\tabcolsep=3pt \fontsize{9}{12}\selectfont
\begin{center}
\textbf{Table 2.}
\textit{Type I error and power with the null hypothesis
$(\beta_0,~\beta_1,~\beta_2)=(1,~1,~0)$}\label{tab2}
\vspace{10pt}

\begin{tabular}{ccc| cc|cc|cc }
 \hline
 &\multicolumn{2}{c|}{$\log(\epsilon)\sim Unif$(-2,2)}&\multicolumn{2}{c|}{$ \log(\epsilon) \sim N(0,1)$ }&\multicolumn{2}{c|}{$\epsilon \sim f_1(\cdot)$}&\multicolumn{2}{c}{$\epsilon \sim f_2(\cdot)$}\\
 \cline{2-3} \cline{4-5} \cline{6-7} \cline{8-9}
$\beta$&$\alpha=0.05$&$\alpha=0.01$& $\alpha=0.05$&$\alpha=0.01$&$\alpha=0.05$&$\alpha=0.01$&$\alpha=0.05$&$\alpha=0.01$\\
\hline
(1.0,~1.0,~0.0)&0.053&0.015&0.049&0.009&0.057&0.013&0.053&0.006\\
(1.0,~1.0,~0.1)&0.338&0.157&0.280&0.120&0.593&0.350&0.772&0.569\\
(1.0,~1.0,~0.2)&0.823&0.637&0.745&0.555&0.987&0.965&0.999&0.997\\
(1.0,~1.0,~0.3)&0.984&0.955&0.980&0.921&1&1&1&1\\
(1.0,~1.0,~0.4)&1&0.999&0.998&0.991&1&1&1&1\\
\hline
\multicolumn{9}{l}{{ $\alpha$ represents the nominal significance level}.}
\end{tabular}
\end{center}
\end{table}

\begin{table}[h!]
\tabcolsep=3pt \fontsize{9}{12}\selectfont
\begin{center}
\textbf{Table 3.}
\textit{Type I error and power with the null hypothesis
$(\beta_0,~\beta_1,~\beta_2)=(1,~0,~0)$}\label{tab3}
\vspace{10pt}

\begin{tabular}{ccc| cc|cc| cc }
 \hline
 &\multicolumn{2}{c|}{$\log(\epsilon)\sim Unif$(-2,2)}&\multicolumn{2}{c|}{$ \log(\epsilon) \sim N(0,1)$ }&\multicolumn{2}{c|}{$\epsilon \sim f_1(\cdot)$}&\multicolumn{2}{c}{$\epsilon \sim f_2(\cdot)$}\\
 \cline{2-3} \cline{4-5} \cline{6-7} \cline{8-9}
$\beta$&$\alpha=0.05$&$\alpha=0.01$& $\alpha=0.05$&$\alpha=0.01$&$\alpha=0.05$&$\alpha=0.01$&$\alpha=0.05$&$\alpha=0.01$\\
\hline
(1.0,~0.0,~0.0)&0.045&0.013&0.057&0.015&0.048&0.008&0.058&0.008\\
(1.0,~0.1,~0.0)&0.270&0.103&0.222&0.084&0.524&0.281&0.683&0.463\\
(1.0,~0.1,~0.1)&0.462&0.258&0.391&0.214&0.809&0.607&0.941&0.828\\
(1.0,~0.2,~0.0)&0.773&0.568&0.663&0.455&0.984&0.933&0.997&0.990\\
(1.0,~0.2,~0.2)&0.965&0.900&0.914&0.810&1&1&1&1\\
\hline
\end{tabular}
\end{center}
\end{table}

\begin{table}[h!]
\tabcolsep=3pt \fontsize{9}{12}\selectfont
\begin{center}
\textbf{Table 4.} \textit{Analysis of the body fat data with LPRE, LARE, LS and LAD }
 \end{center}
{
\begin{center}
 \begin{tabular}{cccc cccc cccc}
 \hline
&\multicolumn{2}{c}{LPRE}&& \multicolumn{2}{c}{LARE}&&\multicolumn{2}{c}{LS}&&\multicolumn{2}{c}{LAD}\\
\cline{2-3}  \cline{5-6}   \cline{8-9} \cline{11-12}
 & Est& $p$-value && Est & $p$-value &&Est   & $p$-value &&Est & $p$-value \\
\hline
$\beta_0$&2.823 (0.026)&0.000&&2.851 (0.027)&0.000&&2.835 (0.022)&0.000&&2.883 (0.029)&0.000\\
$\beta_1$&0.085 (0.038)&0.013&&0.052 (0.027)&0.027&&0.072 (0.031)&0.011&&0.055 (0.038)&0.074\\
$\beta_2$&-0.155 (0.068)&0.011&&-0.205 (0.073)&0.002&&-0.156 (0.056)&0.003&&-0.211 (0.088)&0.008\\
$\beta_3$&-0.103 (0.052)&0.024&&-0.064 (0.044)&0.073&&-0.102 (0.043)&0.009&&-0.064 (0.047)&0.087\\
$\beta_4$&-0.167 (0.076)&0.014&&-0.168 (0.063)&0.004&&-0.134 (0.063)&0.017&&-0.093 (0.081)&0.125\\
$\beta_5$&0.582 (0.091)&0.000&&0.547 (0.084)&0.000&&0.558 (0.075)&0.000&&0.501 (0.105)&0.000\\
$\beta_6$&-0.231 (0.085)&0.003&&-0.200 (0.069)&0.002&&-0.217 (0.07)&0.001&&-0.235 (0.079)&0.001\\
$\beta_7$&0.105 (0.077)&0.086&&0.047 (0.055)&0.196&&0.090 (0.063)&0.076&&0.064 (0.073)&0.190\\
$\beta_8$&0.026 (0.054)&0.315&&0.003 (0.038)&0.469&&0.020 (0.044)&0.327&&-0.011 (0.048)&0.409\\
$\beta_9$&-0.009 (0.036)&0.401&&-0.018 (0.037)&0.313&&-0.005 (0.029)&0.437&&-0.002 (0.032)&0.475\\
$\beta_{10}$&0.081 (0.049)&0.049&&0.034 (0.049)&0.244&&0.051 (0.041)&0.106&&0.008 (0.056)&0.443\\
$\beta_{11}$&0.033 (0.040)&0.205&&0.035 (0.034)&0.152&&0.033 (0.033)&0.157&&0.028 (0.042)&0.252\\
$\beta_{12}$&-0.088 (0.047)&0.031&&-0.084 (0.036)&0.010&&-0.088 (0.039)&0.012&&-0.095 (0.044)&0.015\\
 \hline
  \multicolumn{12}{l}{Est,parameter estimate. The estimated standard deviations are given in the parentheses. }
 \end{tabular}
\end{center}
}
\end{table}

\begin{table}[h!]
\tabcolsep=3pt \fontsize{9}{12}\selectfont
\begin{center}
\textbf{Table 5.} \textit{Comparisons of median prediction errors with LPRE, LARE, LS and LAD }
 \end{center}
{
\begin{center}
 \begin{tabular}{ccccc}
 \hline
 & LPRE & LARE & LS &LAD\\
\hline
 MPE&3.679&  3.957&  3.861&3.933\\
 MPPE&0.039&  0.046&  0.043&0.041\\
 MPAE&0.401&  0.433&  0.418&0.410\\
 MSPE&13.537&  15.660&  14.907&15.468\\
\hline
 \end{tabular}
\end{center}
}
\end{table}

 \end{document}